\documentclass[twocolumn,aps,pra,superscriptaddress,floatfix]{revtex4-2}
\usepackage[utf8]{inputenc}
\usepackage{amsmath}
\usepackage{amssymb}
\usepackage{graphicx}
\usepackage{xcolor}
\usepackage[colorlinks=true,linkcolor=blue,citecolor=blue,urlcolor=blue]{hyperref}

\makeatletter

\usepackage{bbold}

\makeatother

\begin{document}
\title{Phase-sensitive quantum-state texture during thermalization of a
Dzyaloshinskii-Moriya Heisenberg dimer}
\author{Onofre Rojas}
\affiliation{Department of Physics, Institute of Natural Science, Federal University of Lavras, Lavras, MG, Brazil}
\author{Zhirayr Adamyan}
\affiliation{Laboratory of Theoretical Physics, Yerevan State University, 1 Alex Manoogian Str., 0025 Yerevan, Armenia}
\affiliation{CANDLE, Synchrotron Research Institute, 31 Acharyan Str., 0040 Yerevan, Armenia}
\author{Fadwa Benabdallah}
\affiliation{LHEP-MS, Faculty of Sciences, Mohammed V University in Rabat, Morocco}
\author{Vadim Ohanyan}
\affiliation{Laboratory of Theoretical Physics, Yerevan State University, 1 Alex Manoogian Str., 0025 Yerevan, Armenia}
\affiliation{CANDLE, Synchrotron Research Institute, 31 Acharyan Str., 0040 Yerevan, Armenia}

\begin{abstract}
We investigate quantum-state texture during the thermal relaxation of an
anisotropic spin-$1/2$ Heisenberg dimer subject to a longitudinal magnetic
field and a Dzyaloshinskii-Moriya (DM) interaction. Starting from local
transverse and longitudinal couplings to a bosonic bath, we construct a
global weak-coupling Lindblad equation, identify its transition channels,
and discuss the applicability and limitations of the secular approximation. The Gibbs
state is obtained explicitly in the local spin basis, revealing that the DM
coupling controls both the modulus and phase of the single-excitation
coherence by modifying the energy splitting and rotating the chiral phase.
For heating and cooling protocols, we compare rugosity,
fidelity-based texture, and trace-distance texture with coherence, vector
spin chirality, and purity. The relaxation dynamics separates population
thermalization from the decay and precession of deliberately prepared
energy-basis coherence. Their combined contribution to local coherence
produces DM-dependent texture overshoots and, during cooling, damped
oscillations and sign changes of the coherence quadratures. These
transients differ from the evolution of coherence magnitude and purity,
while trace-distance texture weights them differently from the
overlap-based quantifiers. Calculations at reduced dissipation preserve
the texture maxima within the secular model. The results identify a phase-sensitive diagnostic of relaxation.
\end{abstract}
\maketitle

\section{Introduction}

Quantum resource theories provide a systematic language for identifying
which features of a quantum state can enable an operational advantage.
Within this framework, coherence is defined relative to a preferred basis,
whereas imaginarity isolates the resource carried by genuinely complex
density-matrix elements \citep{Streltsov2017,Wu2021}. These notions do not,
however, exhaust the information contained in the arrangement of matrix
elements. States with the same coherence magnitude or purity may display
different amplitude and phase patterns in a fixed laboratory basis.

Quantum-state texture was introduced to quantify precisely this structural
aspect of a density matrix \citep{Parisio2024}. Its original monotone, the
state rugosity, is determined by the overlap with the uniform coherent state
associated with the selected basis and admits a direct measurement protocol.
Recent developments include distance and fidelity-based quantifiers,
entropic and geometric measures, texture witnesses, generalized reference
states, block texture, and connections with purity and criticality
\citep{Wang2025,Zhang2025,Cao2026,Muthuganesan2026,Chen2026,
Greenwood2026,Chu2026,Patra2026,Celeri2026}. These results make clear that
texture is basis dependent and phase sensitive. It is related to, but is not
interchangeable with, coherence, imaginarity, or mixedness.

The behavior of texture in open quantum systems is much less understood.
Environmental coupling redistributes populations and damps coherences,
while coherent Hamiltonian evolution changes their phases; together
these processes can generate texture transients that are not characterized
by the stationary state alone. Recent studies
have shown that texture may respond nontrivially to accelerated motion,
field-induced relaxation, and relativistic transformations
\citep{Huang2025,Huang2026}. Open spin systems also show that coherence,
Fisher-information-based indicators, entanglement, and related correlations
can respond differently to Markovian, non-Markovian, and thermal noise
\citep{Carrion2025,Benabdallah2025}.

Nevertheless, the thermalization of texture in an interacting spin system,
and especially the role of a Hamiltonian phase that survives in local-basis
coherences, remain open problems. This setting also requires care in the
dissipative description. For interacting subsystems, a global master equation
resolves transitions of the full Hamiltonian and provides the natural
weak-coupling route to Gibbs thermalization. The choice between global and
local generators, and between full and partial secularization, can affect
the dynamics and stationary behavior of coupled qubits
\citep{Cattaneo2019,Winczewski2024}. Global treatments have further shown
that nonlocal dissipative channels can qualitatively affect steady-state
coherence, thermometry, and heat transport
\citep{Upadhyay2021,Ullah2023}.

The anisotropic spin-$1/2$ Heisenberg dimer with a Dzyaloshinskii-Moriya
(DM) interaction is a minimal model in which these questions can be addressed
analytically. We consider a longitudinal DM vector, parallel to the
applied magnetic field. Exchange and magnetic field determine the population structure,
temperature controls the mixedness, and the DM interaction modifies both the
energy splitting and the phase of the single-excitation eigenstates. DM-coupled
two-spin systems have been used to study heat rectification and coherence
effects in quantum thermal transport \citep{Upadhyay2021}, while intradimer
DM interactions also appear in quantum dimer magnets displaying thermal Hall
responses \citep{Esaki2024}. More broadly, DM-type spin dynamics is connected
with recent developments in quantum spin dynamics \citep{Liu2024}.

Although a local rotation can remove the DM phase from the isolated
Hamiltonian, that rotation also changes the reference basis. The phase is
therefore physically relevant for texture evaluated in the fixed local spin
basis. The dimer thus offers a transparent setting in which population
relaxation, coherence damping, and phase-sensitive texture can be separated.

Here we investigate the equilibrium and relaxation dynamics of
quantum-state texture in this dimer. We first derive a global thermal
Lindblad equation from the spectral components of local transverse and
longitudinal spin couplings, identify the available Bohr-frequency channels,
and state the regime in which the secular construction is valid. We then
obtain the Gibbs state explicitly in the local basis, thereby separating the
two roles of the DM coupling: it changes both the magnitude and the phase of
the single-excitation coherence.

Finally, for heating and cooling protocols, we compare rugosity, a
fidelity-based measure, and trace-distance texture with the $\ell_1$-norm of
coherence, vector spin chirality, and purity. We show analytically that
rugosity and the fidelity-based texture are monotonic functions of the same
overlap, whereas trace distance depends on the full operator difference and
is not determined by the single reference overlap.
The resulting dynamics exposes texture as a phase-sensitive diagnostic of
thermal relaxation rather than a proxy for coherence magnitude or global
mixedness.

\section{Heisenberg model Hamiltonian}\label{sec:model}

We consider the spin-$1/2$ Heisenberg dimer Hamiltonian
\begin{alignat}{1}
\mathbf{H}= & -J(S^{x}_{1}S^{x}_{2}+S^{y}_{1}S^{y}_{2})-J_{z}S^{z}_{1}S^{z}_{2}\nonumber \\
 & -h(S^{z}_{1}+S^{z}_{2})-D(S^{x}_{1}S^{y}_{2}-S^{y}_{1}S^{x}_{2}),\label{eq:Hamiltonian}
\end{alignat}
where $\boldsymbol{S}_{j}=(S^{x}_{j},S^{y}_{j},S^{z}_{j})$ are the
spin-1/2 operators, $J$ is the $XY$ exchange coupling, $J_{z}$
the longitudinal exchange coupling, $h$ is an external magnetic field
along the $z$-axis, and $D$ is the strength of the Dzyaloshinskii-Moriya
(DM) interaction \citep{Dzyaloshinsky1958,Moriya1960}.
The sign of the DM term is conventional because the spectrum depends
on $D^2$; we retain it to obtain the KNB relation $D=-E\sin\vartheta$.

It is convenient to work in the energy eigenbasis $\{\vert\varphi_{1}\rangle,\vert\varphi_{2}\rangle,\vert\varphi_{3}\rangle,\vert\varphi_{4}\rangle\}$,
where the eigenstates are identified as
\begin{alignat}{1}
\vert\varphi_{1}\rangle= & \vert\uparrow\uparrow\rangle,\\
\vert\varphi_{2}\rangle= & \tfrac{\vert\uparrow\downarrow\rangle+{\rm e}^{-i\phi}\vert\downarrow\uparrow\rangle}{\sqrt{2}},\\
\vert\varphi_{3}\rangle= & \tfrac{\vert\uparrow\downarrow\rangle-{\rm e}^{-i\phi}\vert\downarrow\uparrow\rangle}{\sqrt{2}},\\
\vert\varphi_{4}\rangle= & \vert\downarrow\downarrow\rangle,
\end{alignat}
where the DM-induced phase is $\phi=\arg(J+iD)$. The states $\vert\varphi_{2}\rangle$
and $\vert\varphi_{3}\rangle$ have zero total magnetization ($M=0$),
while $\vert\varphi_{1}\rangle$ and $\vert\varphi_{4}\rangle$ correspond
to full polarization along $+z$ and $-z$, respectively.

The corresponding energy eigenvalues are
\begin{alignat}{1}
\varepsilon_{1} & =-\frac{J_{z}}{4}-h,\\
\varepsilon_{2} & =\frac{J_{z}}{4}-\frac{K}{2},\\
\varepsilon_{3} & =\frac{J_{z}}{4}+\frac{K}{2},\\
\varepsilon_{4} & =-\frac{J_{z}}{4}+h,\label{eq:energies}
\end{alignat}
where $K=\sqrt{J^{2}+D^{2}}$.
We note that the eigenstates fall into well-defined magnetization
sectors, with the subspace spanned by $\vert\varphi_{2}\rangle$ and
$\vert\varphi_{3}\rangle$ forming entangled Bell-like states whose relative
phase is twisted by the DM interaction. Their splitting is governed by
the parameter $K$, which treats the XY and
DM couplings on equal footing and is positive except at $J=D=0$. The external field $h$ splits the
polarized states, whereas for $h=0$ the two polarized levels are
degenerate. More generally, the spectrum
is invariant under $h\to-h$, which interchanges
$\varepsilon_1$ and $\varepsilon_4$.

The DM term can also arise from the Katsura-Nagaosa-Balatsky (KNB)
mechanism of spin-induced ferroelectricity \cite{KNB1}. This microscopic
mechanism of magnetoelectric coupling relates the spin-induced polarization
of certain magnetic materials to a local spin current. The resulting dipole
moment on the bond connecting sites $i$ and $j$ is
\begin{eqnarray}
\boldsymbol{\sf P}_{ij}=g_{ij} \mathbf{e}_{ij}\times\left(\mathbf{S}_i\times\mathbf{S}_j\right),
\end{eqnarray}
where $\mathbf{e}_{ij}$ is the unit vector directed from site $i$ to
site $j$, and $\mathbf{S}_i$ and $\mathbf{S}_j$ are the corresponding spin
operators. The bond-dependent coefficient $g_{ij}$ reflects the local
quantum-chemical environment \cite{KNB1,KNB2}. Electric-field effects in
quantum spin systems governed by the KNB mechanism have been explored in a
variety of settings \cite{bro13,thakur18,XYZ,oha20,mench15,bar21,sznajd18,
sznajd19,baran18,sc_KNB,JJJ,dis,ent_tran,bre1}. For a spin dimer whose bond
defines the $x$ axis ($\mathbf{e}_{ij}=\mathbf{e}_x$), the KNB polarization
becomes
\begin{eqnarray}
&&{\sf P}_x=0, \\
&&{\sf P_y}=g\left(S_1^yS_2^x-S_1^xS_2^y\right), \nonumber\\
&&{\sf P_z}=g\left(S_1^zS_2^x-S_1^xS_2^z\right). \nonumber
\end{eqnarray}
If the electric field lies in the $(x,y)$ plane, its interaction with
the KNB polarization, $-\mathbf{E}\cdot\boldsymbol{\sf P}$, contributes the
following term to the spin Hamiltonian:
\begin{eqnarray}
H_{KNB}=E_y(S_1^xS_2^y-S_1^yS_2^x).
\end{eqnarray}
The DM coefficient in Eq.~\eqref{eq:Hamiltonian} can therefore be
related to the applied electric field through
\begin{eqnarray}
D=-E\sin\vartheta
\end{eqnarray}
where the material coefficient $g$ has been absorbed into the effective
field magnitude $E$, and $\vartheta$ is the angle between the electric field
and the dimer axis. The DM coupling can thus be controlled either by changing
the field magnitude or by rotating the field \cite{ent_tran}.

\subsection{Thermal state in the local basis}

At inverse temperature $\beta=(k_{B}T)^{-1}$, the equilibrium state
is described by the Gibbs density matrix
\begin{equation}
\rho_{\beta}=\frac{e^{-\beta\mathbf{H}}}{Z}
=\sum_{m=1}^{4}p_{m}\vert\varphi_{m}\rangle\langle\varphi_{m}\vert,
\quad
p_{m}=\frac{e^{-\beta\varepsilon_{m}}}{Z},
\end{equation}
where
\begin{align}
Z={}&2e^{\beta J_{z}/4}\cosh(\beta h)
+2e^{-\beta J_{z}/4}\cosh(\beta K/2).
\end{align}
In the local computational basis
$\{\vert\uparrow\uparrow\rangle,\vert\uparrow\downarrow\rangle,
\vert\downarrow\uparrow\rangle,\vert\downarrow\downarrow\rangle\}$,
the Gibbs state takes the form
\begin{equation}
\varrho_{\beta}=
\begin{pmatrix}
p_{1} & 0 & 0 & 0\\
0 & \frac{p_{2}+p_{3}}{2}
& \frac{p_{2}-p_{3}}{2}e^{i\phi} & 0\\
0 & \frac{p_{2}-p_{3}}{2}e^{-i\phi}
& \frac{p_{2}+p_{3}}{2} & 0\\
0 & 0 & 0 & p_{4}
\end{pmatrix}.
\label{eq:thermal-local}
\end{equation}
The DM interaction changes the coherence magnitude through the
$K$-dependent population difference $p_{2}-p_{3}$ and rotates its phase
through $\phi$. The latter effect is explicitly basis dependent: a local
$z$ rotation can remove the DM phase from the Hamiltonian \eqref{eq:Hamiltonian}, but it also
rotates the reference basis used to define texture. We therefore keep
the laboratory spin basis fixed throughout the analysis.

\subsection{Bohr frequencies}

For any pair of energy eigenstates $\vert\varphi_{m}\rangle,\vert\varphi_{n}\rangle$,
we use
\begin{equation}
\omega_{mn}=\varepsilon_m-\varepsilon_n,\qquad
\Omega_{mn}=|\omega_{mn}|.
\end{equation}
Thus $\omega_{mn}$ is the signed Bohr frequency, useful for coherent
phase factors, whereas $\Omega_{mn}$ is the positive transition gap
at which the bath functions are evaluated.
The direction of energy flow is incorporated in the Lindblad operators
themselves: energy-lowering operators describe emission processes,
whereas their adjoints describe absorption processes.

\begin{figure}
\centering
\includegraphics[scale=0.5]{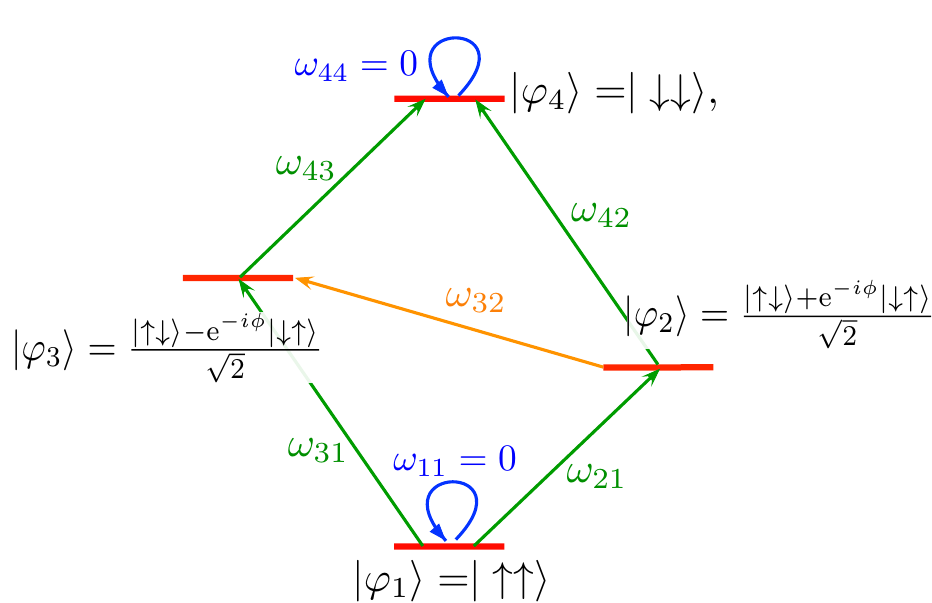}
\caption{Schematic representation of the transverse and longitudinal transitions.
The links are labeled by the signed gaps $\omega_{21},\omega_{31},\omega_{42},\omega_{43}$.
The physical absorption direction on each link is determined by the actual energy ordering.
Away from level crossings, the loop at $\omega=0$ denotes pure dephasing.
Since $\langle\varphi_1|S^{z}|\varphi_1\rangle=+1/2$ and
$\langle\varphi_4|S^{z}|\varphi_4\rangle=-1/2$, this zero-frequency
channel affects both polarized states, but with opposite signs in the
corresponding matrix elements.}
\label{fig:schm-bohr}
\end{figure}

The local raising and lowering operators on sites $j=1,2$ are
$S^{\pm}_{1}=S^{\pm}\otimes\mathbb{1}$ and
$S^{\pm}_{2}=\mathbb{1}\otimes S^{\pm}$, which change the total
magnetization by $\Delta M=\pm1$. These operators generate the
transition chain schematically illustrated in Fig.~\ref{fig:schm-bohr}.
Besides the finite-frequency channels indicated by arrows, the figure
also displays a loop at $\omega=0$, corresponding away from level
crossings to pure dephasing processes acting on the polarized states.

For completeness, the six pairwise gaps are
\begin{alignat}{2}
\omega_{21} & =\varepsilon_{2}-\varepsilon_{1} &  & =h+\tfrac{J_{z}}{2}-\tfrac{K}{2}, \nonumber \\
\omega_{31} & =\varepsilon_{3}-\varepsilon_{1} &  & =h+\tfrac{J_{z}}{2}+\tfrac{K}{2},\nonumber \\
\omega_{42} & =\varepsilon_{4}-\varepsilon_{2} &  & =h-\tfrac{J_{z}}{2}+\tfrac{K}{2},\nonumber \\
\omega_{43} & =\varepsilon_{4}-\varepsilon_{3} &  & =h-\tfrac{J_{z}}{2}-\tfrac{K}{2},\nonumber \\
\omega_{32} & =\varepsilon_{3}-\varepsilon_{2} &  & =K,\nonumber \\
\omega_{41} & =\varepsilon_{4}-\varepsilon_{1} &  & =2h.\label{eq:bohr}
\end{alignat}

The gap
$\omega_{41}$ is not an active channel here because the chosen local
couplings have no $|\varphi_1\rangle\leftrightarrow|\varphi_4\rangle$
matrix element; similarly, $|\varphi_{2,3}\rangle$ have no diagonal
$S_j^z$ contribution to nondegenerate zero-frequency dephasing.

\subsection{Transverse spectral components}

In the eigenstate basis, the local spin-lowering operators decompose into spectral
components connecting states separated by a single Bohr frequency. In
general, $S_j^- =\sum_{m,n}\langle\varphi_m|S_j^-|\varphi_n\rangle
|\varphi_m\rangle\langle\varphi_n|$; retaining its nonzero matrix elements
gives the components below.

We label a transition operator by its final and initial states,
$\mathbf{A}^{mn}_{j}\propto\vert\varphi_m\rangle
\langle\varphi_n\vert$. The nonzero components of $S_j^{-}$ read

\begin{align}
\mathbf{A}^{21}_{j}
&=\tfrac{{\rm e}^{i\phi(2-j)}}{\sqrt{2}}
\vert\varphi_{2}\rangle\langle\varphi_{1}\vert,\\
\mathbf{A}^{31}_{j}
&=\tfrac{(-1)^{j}{\rm e}^{i\phi(2-j)}}{\sqrt{2}}
\vert\varphi_{3}\rangle\langle\varphi_{1}\vert,
\label{eq:A1top}\\
\mathbf{A}^{42}_{j}
&=\tfrac{{\rm e}^{-i\phi(j-1)}}{\sqrt{2}}
\vert\varphi_{4}\rangle\langle\varphi_{2}\vert,\\
\mathbf{A}^{43}_{j}
&=\tfrac{(-1)^{j+1}{\rm e}^{-i\phi(j-1)}}{\sqrt{2}}
\vert\varphi_{4}\rangle\langle\varphi_{3}\vert,
\label{eq:A1bottom}
\end{align}
where $j\in\{1,2\}$ labels the spin sites.

The complex amplitudes in Eqs.~\eqref{eq:A1top} and \eqref{eq:A1bottom}
originate from the DM-induced phase in the eigenstates. For uncorrelated
local bath channels, the dissipators associated with $j=1$ and $j=2$ are
added incoherently, so these scalar phases do not by themselves produce
interference in the transition rates. Interference between the two
paths would require correlated bath channels and the corresponding
cross-damping terms.

The spin-raising components follow as adjoints $\mathbf{A}^{nm}_{j}=\big[\mathbf{A}^{mn}_{j}\big]^{\dagger}$ and describe the reverse transitions.

\subsection{Longitudinal spectral components}

For longitudinal (dephasing) coupling, the relevant system operators are $S^{z}_{j}$.
In the energy eigenstate basis $\{\vert\varphi_{1}\rangle,\vert\varphi_{2}\rangle,\vert\varphi_{3}\rangle,\vert\varphi_{4}\rangle\}$,
the nonzero matrix elements are
\begin{align}
\langle\varphi_{1}|S^{z}_{1,2}|\varphi_{1}\rangle &= +\tfrac{1}{2},\label{eq:phi_1S^z}\\
\langle\varphi_{4}|S^{z}_{1,2}|\varphi_{4}\rangle &= -\tfrac{1}{2},\\
\langle\varphi_{2}|S^{z}_{1,2}|\varphi_{2}\rangle &= \langle\varphi_{3}|S^{z}_{1,2}|\varphi_{3}\rangle = 0,\\
\langle\varphi_{3}|S^{z}_{1}|\varphi_{2}\rangle &= -\langle\varphi_{3}|S^{z}_{2}|\varphi_{2}\rangle = +\tfrac{1}{2},\label{eq:phi_32S^z}
\end{align}
The opposite signs for sites $1$ and $2$ arise because $S^{z}_{1}$ and
$S^{z}_{2}$ act with opposite signs on $\vert\uparrow\downarrow\rangle$
and $\vert\downarrow\uparrow\rangle$.

As a consequence, longitudinal couplings yield two types of spectral
components:

(i) Away from level crossings, the zero-frequency terms correspond to
pure dephasing, which randomizes relative phases without energy
exchange. For the $S_j^z$ coupling they are
\begin{equation}
\mathbf{A}^{0}_{j}= \tfrac{1}{2}\,\vert\varphi_{1}\rangle\langle\varphi_{1}\vert - \tfrac{1}{2}\,\vert\varphi_{4}\rangle\langle\varphi_{4}\vert,
\end{equation}
with no diagonal contribution on $\vert\varphi_{2,3}\rangle$ since their
$S^{z}$ expectation values vanish.

(ii) Finite-gap terms between levels $2$ and $3$, separated by
$\omega_{32}=\varepsilon_{3}-\varepsilon_{2}=K$,
which mix states $\vert\varphi_{2}\rangle$ and $\vert\varphi_{3}\rangle$.

With the final-initial labeling convention, the upward transition
operator is
\begin{alignat}{1}
\mathbf{A}^{32}_{j}= & \tfrac{(-1)^{j+1}}{2}\,\vert\varphi_{3}\rangle\langle\varphi_{2}\vert,
\end{alignat}
and its adjoint is the energy-lowering operator,
\begin{equation}
\mathbf{A}^{23}_{j}=[\mathbf{A}^{32}_{j}]^{\dagger}=\tfrac{(-1)^{j+1}}{2}\,\vert\varphi_{2}\rangle\langle\varphi_{3}\vert.
\end{equation}

These operators complete the longitudinal decomposition and play a
central role in the dissipative dynamics discussed in Sec.~\ref{sec:master}.

\section{Dimer-environment interaction}\label{sec:master}

We describe the dissipative dynamics of the Heisenberg dimer coupled to a
thermal reservoir using the Lindblad formalism, which guarantees
completely positive, trace-preserving, and Markovian evolution. In
the secular (global) approach, one first diagonalizes the Hamiltonian
\eqref{eq:Hamiltonian} and expresses the system-bath coupling operators in terms
of spectral components $\mathbf{A}^{mn}_{j}$ that connect eigenstates
$\vert\varphi_{m}\rangle$ and $\vert\varphi_{n}\rangle$ with Bohr frequency
$\omega_{mn}=\varepsilon_{m}-\varepsilon_{n}$
\citep{lindblad,BreuerPetruccione}.

The relevant components are the transverse channels connecting
$1\leftrightarrow2$, $1\leftrightarrow3$, $2\leftrightarrow4$, and
$3\leftrightarrow4$, together with the longitudinal channels
$2\leftrightarrow3$ and the zero-frequency dephasing component.

For definiteness, we consider a single bosonic thermal environment
represented by local Hermitian coupling channels,
\begin{equation}
\mathbf{H}_{\mathrm{I}}
=\sum_{j=1}^{2}\sum_{\mu}
\mathbf{S}_{j}^{\mu}\otimes\mathbf{B}_{j}^{\mu},
\quad
\mu\in\{x,z\},
\label{eq:interaction}
\end{equation}
at inverse temperature $\beta$. We retain $x$ and $z$ as minimal
transverse and longitudinal channels; for an isotropic transverse bath,
$S^y$ would introduce no new transition class.
We assume that the bath operators
associated with different local channels are uncorrelated and have the
same autocorrelation spectrum,
\begin{equation}
\langle \mathbf{B}_{j}^{\mu}(t)\mathbf{B}_{k}^{\nu}(0)\rangle
=\delta_{jk}\delta_{\mu\nu}C(t).
\label{eq:bath-correlation}
\end{equation}
The Markov approximation assumes that the bath correlation time is
short compared with the system relaxation time; it does not require
$C(t)=\delta(t)$ literally.
The uncorrelated local channels in Eq.~\eqref{eq:bath-correlation}
eliminate cross-damping terms, and their common autocorrelation spectrum
defines the spectral density $\kappa(\omega)$. This is an assumption
of independent local noise channels at a common temperature, rather than
collective coupling to a common bath operator; a shared temperature alone
does not imply the absence of bath correlations.
The
$S_j^x$ channel contains the raising and lowering matrix elements
listed above. Since $S_j^x=(S_j^++S_j^-)/2$, its spectral components
carry an additional factor $1/2$ relative to Eqs.
\eqref{eq:A1top} and \eqref{eq:A1bottom}. This factor is retained in
$\mathbf{S}_{j}^{x}(\omega)$ rather than absorbed into
$\kappa(\omega)$. The operator $S_j^z$ generates the longitudinal
channels.

The corresponding Lindblad master equation, with all transitions
sharing the same Bohr frequency grouped into the same spectral
component, reads
\begin{equation}
\dot{\boldsymbol{\rho}}=-i[\mathbf{H},\boldsymbol{\rho}]+\mathcal{L}[\boldsymbol{\rho}],
\end{equation}
where the dissipator is written in the standard global form
\begin{align}
\mathcal{L}[\boldsymbol{\rho}]
=\sum_{j=1}^{2}\sum_{\mu\in\{x,z\}}
\sum_{\omega>0}\bigl\{&
\gamma^{-}(\omega)
\mathcal{D}[\mathbf{S}_{j}^{\mu}(\omega)]\boldsymbol{\rho}\nonumber\\
&+\gamma^{+}(\omega)
\mathcal{D}[(\mathbf{S}_{j}^{\mu}(\omega))^{\dagger}]
\boldsymbol{\rho}\bigr\}\nonumber\\
&+\gamma_{0}\sum_{j=1}^{2}
\mathcal{D}[\mathbf{A}_{j}^{0}]\boldsymbol{\rho},
\label{eq:Mast-eq}
\end{align}

with
$\gamma^-(\omega)$, $\gamma^+(\omega)$, and $\gamma_0$ being the emission,
absorption, and zero-frequency dephasing coefficients, respectively.

Here the dissipator given by
\begin{equation}
\mathcal{D}[\mathbf{L}]\boldsymbol{\rho}=\mathbf{L}\boldsymbol{\rho}\mathbf{L}^{\dagger}-\tfrac{1}{2}\{\mathbf{L}^{\dagger}\mathbf{L},\boldsymbol{\rho}\}.
\end{equation}

This is the Gorini-Kossakowski-Sudarshan-Lindblad structure for a
time-homogeneous Markovian quantum dynamical semigroup
\citep{Gorini1976,lindblad,BreuerPetruccione}.
For a system coupling operator $\mathbf{S}_{j}^{\mu}$, its positive-frequency
component is defined by
\begin{equation}
\mathbf{S}_{j}^{\mu}(\omega)=
\sum_{\varepsilon_{n}-\varepsilon_{m}=\omega}
\vert\varphi_{m}\rangle\langle\varphi_{m}\vert
\mathbf{S}_{j}^{\mu}
\vert\varphi_{n}\rangle\langle\varphi_{n}\vert ,
\qquad \omega>0.
\end{equation}
Thus $\mathbf{S}_{j}^{\mu}(\omega)$ always lowers the system energy by
$\omega$, independently of the labels $m$ and $n$. Its adjoint raises
the energy. For a nondegenerate Hamiltonian, the zero-frequency
component is diagonal in the energy basis and produces pure dephasing.
A possible
Lamb-shift Hamiltonian has been omitted, or equivalently absorbed into
a renormalized $\mathbf{H}$. The operators $\mathbf{A}_{j}^{mn}$
introduced above identify the explicit rank-one transitions. For a
transverse pair, $\mathbf{S}_{j}^{x}(\omega)$ is one half of the member
of $\{\mathbf{A}_{j}^{mn},\mathbf{A}_{j}^{nm}\}$ that lowers the
energy. For the longitudinal channel,
$\mathbf{S}_{j}^{z}(K)=\mathbf{A}_{j}^{23}$ and
$\mathbf{S}_{j}^{z}(0)=\mathbf{A}_{j}^{0}$.

At an exact level crossing,
two or more energy eigenstates belong to the same zero-frequency sector.
In that case, $\mathbf{S}_j^\mu(0)$ must be reconstructed by projecting
$\mathbf{S}_j^\mu$ onto the full degenerate subspace. The resulting
zero-frequency jump operators need not be purely diagonal and may also
mix populations inside that subspace. Therefore, the simple nondegenerate
form of Eq.~\eqref{eq:Mast-eq} is not used exactly at such crossings.

\subsection{Bath spectral density and detailed balance}

For a bosonic reservoir, the downward and upward coefficients are
\begin{alignat}{1}
\gamma^{-}(\omega) & =\kappa(\omega)[1+n(\omega)],
\label{eq:gm-m}\\
\gamma^{+}(\omega) & =\kappa(\omega)n(\omega),
\label{eq:gm-p}
\end{alignat}
where
\begin{equation}
n(\omega)=\frac{1}{e^{\beta\omega}-1}
\end{equation}
is the Bose-Einstein distribution at inverse temperature
$\beta=1/(k_{B}T)$.  Here $\kappa(\omega)$ is the positive-frequency spectral density
associated with $C(t)$ and determines the frequency dependence of the rates.

For a specific transition pair with
$\Omega_{mn}=|\varepsilon_m-\varepsilon_n|>0$, we introduce the
abbreviations
\begin{align}
\gamma_{mn}^{-}
&\equiv\gamma^{-}(\Omega_{mn})
=\kappa(\Omega_{mn})[1+n(\Omega_{mn})],
\nonumber\\
\gamma_{mn}^{+}
&\equiv\gamma^{+}(\Omega_{mn})
=\kappa(\Omega_{mn})n(\Omega_{mn}).
\label{eq:pair-rates}
\end{align}
Thus the subscripts $m,n$ identify the transition gap at which the
common bath rate is evaluated; they do not label different reservoirs.
Since $\Omega_{mn}=\Omega_{nm}$, one also has
$\gamma_{mn}^{\pm}=\gamma_{nm}^{\pm}$.

Equivalently, the pair rates satisfy detailed balance,
\begin{equation}
\frac{\gamma_{mn}^{+}}{\gamma_{mn}^{-}}
=e^{-\beta\Omega_{mn}},
\end{equation}
which makes the Gibbs state stationary. Relaxation to that state follows
when the nonzero dissipative channels allow population transfer among
all energy levels \citep{BreuerPetruccione}. The dephasing
coefficient is determined by the
zero-frequency limit of the complete noise spectrum. In particular,
for an Ohmic bath one must take the limit of the product containing the
thermal factor; simply substituting $\kappa(0)=0$ would incorrectly
discard thermal dephasing \citep{BreuerPetruccione,Leggett}.
Fermionic and spin baths require different occupation statistics or
correlation functions and lie beyond the present bosonic-bath model.

We use an Ohmic spectral density with an exponential cutoff,

\begin{equation}
\kappa(\omega)=\eta\,\omega\,e^{-\omega/\omega_{c}},
\end{equation}
where $\eta$ sets the dissipation strength and $\omega_{c}$ is the
high-frequency cutoff \citep{CDuan,Barr,Leggett}. Evaluating this spectrum
at the positive Bohr frequencies gives the
upward and downward rates in Eqs.~\eqref{eq:gm-m} and \eqref{eq:gm-p}.

\subsection{Coherences}

Full secularization requires a coarse-graining time $\Delta t$ such that
$\tau_B\ll\Delta t\ll\tau_R$, while
$|\omega-\omega'|\Delta t\gg1$ for distinct frequencies coupled
by the same bath channel. Here $\tau_B$ is the bath correlation time
and $\tau_R$ the relaxation time. Thus, relevant dissipative linewidths
must be small compared with the frequency separations of the discarded
terms. Exact equal-frequency components are retained together; this
grouping does not justify discarding nearby unequal-frequency terms
\citep{Cattaneo2019,Winczewski2024}.

In the full secular approximation, and provided distinct transitions
do not have degenerate or nearly degenerate Bohr frequencies, each
off-diagonal element of the density matrix in the energy basis evolves
independently \citep{BreuerPetruccione,Cattaneo2019,Winczewski2024}:
\begin{equation}
\dot{\rho}_{mn}=-i\omega_{mn}\rho_{mn}-\Gamma_{mn}\rho_{mn},\quad m\ne n.
\end{equation}
Here $\omega_{mn}$ is the signed Bohr frequency of
Eq.~\eqref{eq:bohr}, and the decay rate is
\begin{equation}
\Gamma_{mn}=\tfrac{1}{2}\big(\Lambda_{m}+\Lambda_{n}\big)+\Gamma^{z}_{mn}.
\end{equation}

Here the first contribution originates from the population outflow
rates $\Lambda_m$ and $\Lambda_n$ defined in
Eq.~\eqref{eq:outflow-rates},
while the second term accounts
for additional longitudinal dephasing. The latter can be written as
\begin{equation}
\Gamma^{z}_{mn}=\Gamma^{z}_{nm}=\tfrac{\gamma_{0}}{2}\sum^{2}_{j=1}\big(\langle\varphi_{m}|S^{z}_{j}|\varphi_{m}\rangle\!-
\!\langle\varphi_{n}|S^{z}_{j}|\varphi_{n}\rangle\big)^{2},
\end{equation}
which damps coherences between states with different local magnetizations.
Using Eqs.~\eqref{eq:phi_1S^z}-\eqref{eq:phi_32S^z}, we obtain
\begin{alignat}{1}
\Gamma^{z}_{21} & =\Gamma^{z}_{31}=\Gamma^{z}_{42}=\Gamma^{z}_{43}=\gamma_{0}/4,\\
\Gamma^{z}_{32} & =0,\\
\Gamma^{z}_{41} & =\gamma_{0},
\end{alignat}

Therefore the general solution is

\begin{equation}
\rho_{mn}(t)=\rho_{mn}(0)\;e^{-i\omega_{mn}t}\;e^{-\Gamma_{mn}t}.
\end{equation}

For the frequency degeneracies present in the numerical parameter set,
the grouping prescribed in Eq.~\eqref{eq:Mast-eq} leaves the only dynamically
relevant energy-basis coherence, $\rho_{23}$ (and its conjugate
$\rho_{32}$), governed by the same scalar equation.

Away from degenerate or nearly degenerate Bohr frequencies, and provided
that none of the required dissipative channels is absent, all
energy-basis coherences vanish and the system relaxes to the Gibbs
state. This does not imply the absence of coherence in the local basis;
the thermal coherence in Eq.~\eqref{eq:thermal-local} generally
remains nonzero.

To evaluate the texture quantifiers, we express the density
operator in the local computational basis
$\{\vert\uparrow\uparrow\rangle,\vert\uparrow\downarrow\rangle,\vert\downarrow\uparrow\rangle,\vert\downarrow\downarrow\rangle\}$.
The state $\rho(t)$, initially written in the energy eigenbasis
$\{\vert\varphi_{1}\rangle,\dots,\vert\varphi_{4}\rangle\}$, has the
matrix form:

\begin{equation}
\rho=\left[\begin{array}{cccc}
r_{1} & \rho_{12} & \rho_{13} & \rho_{14}\\
\rho^{*}_{12} & r_{2} & \rho_{23} & \rho_{24}\\
\rho^{*}_{13} & \rho^{*}_{23} & r_{3} & \rho_{34}\\
\rho^{*}_{14} & \rho^{*}_{24} & \rho^{*}_{34} & r_{4}
\end{array}\right].
\end{equation}

To transform this state into the local computational basis, we apply the unitary
matrix ${\bf U}$, constructed from the eigenstates defined above.
The transformation reads:

\begin{equation}
{\bf U}=\tfrac{1}{\sqrt{2}}\left[\begin{array}{cccc}
\sqrt{2} & 0 & 0 & 0\\
0 & 1 & 1 & 0\\
0 & e^{-i\phi} & -e^{-i\phi} & 0\\
0 & 0 & 0 & \sqrt{2}
\end{array}\right],
\end{equation}
where $\phi=\arg(J+iD)$.

The density operator $\varrho$ in the local computational basis is then
given by

\begin{equation}
\varrho={\bf U}\rho {\bf U}^{\dagger}=\left[\begin{array}{cccc}
\varrho_{11} & \varrho_{12} & \varrho_{13} & \varrho_{14}\\
\varrho^{*}_{12} & \varrho_{22} & \varrho_{23} & \varrho_{24}\\
\varrho^{*}_{13} & \varrho^{*}_{23} & \varrho_{33} & \varrho_{34}\\
\varrho^{*}_{14} & \varrho^{*}_{24} & \varrho^{*}_{34} & \varrho_{44}
\end{array}\right].
\end{equation}
The components of $\varrho$ are related to the elements of $\rho$
by

\begin{alignat}{1}
\varrho_{11}(t)= & r_{1}(t),\\
\varrho_{12}(t)= & \tfrac{\left(\rho_{12}(t)+\rho_{13}(t)\right)}{\sqrt{2}},\\
\varrho_{13}(t)= & \tfrac{\left(\rho_{12}(t)-\rho_{13}(t)\right)}{\sqrt{2}}{\rm e}^{i\phi},\\
\varrho_{14}(t)= & \rho_{14}(t),\\
\varrho_{22}(t)= & \tfrac{\left(2\Re(\rho_{23}(t))+r_{2}(t)+r_{3}(t)\right)}{2},\\
\varrho_{23}(t)= & \tfrac{\left(-i2\Im(\rho_{23}(t))+r_{2}(t)-r_{3}(t)\right)}{2}{\rm e}^{i\phi},\label{eq:local-coherence-decomposition}\\
\varrho_{24}(t)= & \tfrac{\left(\rho_{24}(t)+\rho_{34}(t)\right)}{\sqrt{2}},\\
\varrho_{33}(t)= & \tfrac{\left(-2\Re(\rho_{23}(t))+r_{2}(t)+r_{3}(t)\right)}{2},\\
\varrho_{34}(t)= & \tfrac{\left(\rho_{24}(t)-\rho_{34}(t)\right)}{\sqrt{2}}{\rm e}^{-i\phi},\\
\varrho_{44}(t)= & r_{4}(t).
\end{alignat}

We use the transformed density matrix $\varrho(t)$ in the following section
because all texture measures are defined relative to the local computational
basis.

\section{Texture measures}

Quantum-state texture complements quantities such as energy, purity, and
entanglement by resolving basis-dependent density-matrix structure
\citep{Parisio2024,Wang2025}. It is sensitive to coherence, imaginarity,
and interference effects that may not affect populations or entropy.

Since texture is basis dependent, we adopt the natural computational basis $\{\vert\uparrow\uparrow\rangle,\vert\uparrow\downarrow\rangle,\vert\downarrow\uparrow\rangle,\vert\downarrow\downarrow\rangle\}$,
which corresponds to local spin configurations along the magnetic-field
direction.

The degree of texture is then quantified by comparing the system state
with the coherent flat state

\begin{equation}
\vert f_{1}\rangle=\frac{1}{2}(\vert\uparrow\uparrow\rangle+\vert\uparrow\downarrow\rangle+\vert\downarrow\uparrow\rangle+\vert\downarrow\downarrow\rangle).
\end{equation}
The subscript $1$ denotes the equal-phase, equal-amplitude flat
reference state in this basis \citep{Parisio2024,Wang2025}.
The projector $\varrho_{f_1}=\vert f_1\rangle\langle f_1\vert$ is the
textureless reference in this basis \citep{Parisio2024,Wang2025}.

\subsection{Rugosity}

Rugosity quantifies the deviation of a state $\varrho$ from the
textureless projector $\varrho_{f_1}$ and is defined as
\citep{Parisio2024}
\[
R(\varrho)=-\ln\left(\langle f_{1}|\varrho|f_{1}\rangle\right).
\]

\subsection{Fidelity-based texture}

For the pure reference $\varrho_{f_1}$, the squared Uhlmann-fidelity
convention of Ref.~\citep{Wang2025} is

\[
F(\varrho,\varrho_{f_1})=\langle f_{1}|\varrho|f_{1}\rangle.
\]
With this convention, $F=1$ if and only if
$\varrho=\vert f_{1}\rangle\langle f_{1}\vert$. The corresponding
texture quantifier is

\[
T_{F}(\varrho)=1-F(\varrho,\varrho_{f_1})=1-\langle f_{1}|\varrho|f_{1}\rangle.
\]
This quantity is directly measurable as a projection probability. Rugosity and
$T_F$ are not independent measures: both are monotonic functions of
the same overlap, with $R=-\ln(1-T_F)$. Consequently, they always rank
states in the same order, although their numerical scales differ.

\subsection{Trace distance}

For two density matrices $\varrho$ and $\sigma$, the trace distance
is defined as
\[
D_{\text{tr}}(\varrho,\sigma)=\frac{1}{2}\|\varrho-\sigma\|_{1},
\]
where $\|A\|_{1}=\operatorname{Tr}\left(\sqrt{A^{\dagger}A}\right)$ is the
Schatten-1 norm, i.e., the sum of the singular values (or absolute
values of the eigenvalues if $A$ is Hermitian).

In our case, the corresponding texture measure is
\begin{equation}
T_{\text{tr}}(\varrho)=\frac{1}{2}\|\varrho-\varrho_{f_{1}}\|_{1},
\end{equation}
where $\varrho_{f_{1}}=\vert f_{1}\rangle\langle f_{1}\vert$ is the projector
onto the equal-superposition state in the chosen basis
\citep{Wang2025}.
Although \(T_{\rm tr}\) is not determined by the single reference
overlap, it is bounded by the corresponding fidelity-based texture:
\[
T_F\leqslant T_{\rm tr}\leqslant\sqrt{T_F}.
\]

For the thermal state in Eq.~\eqref{eq:thermal-local}, the common overlap
entering rugosity and fidelity is
\begin{equation}
F_{\beta}
=\langle f_{1}|\varrho_{\beta}|f_{1}\rangle
=\frac{1+(p_{2}-p_{3})\cos\phi}{4}.
\label{eq:thermal-overlap}
\end{equation}
Using $\cos\phi=J/K$, this can also be written as
\begin{equation}
F_\beta=\frac{1}{4}\left[
1+\frac{2J}{KZ}e^{-\beta J_z/4}
\sinh\left(\frac{\beta K}{2}\right)\right],
\qquad K>0.
\label{eq:thermal-overlap-explicit}
\end{equation}
This compact result separates the population imbalance governed by
$K$ from the phase rotation governed by $\phi$. The trace distance is
not fixed by this single overlap and may therefore resolve differences
between states with the same rugosity and $T_F$. This is a possibility,
not a claim that the scalar trace-distance texture distinguishes every
pair of distinct thermal states.

It is important to distinguish the fixed local basis used here from
the energy basis. If the texture reference is defined in the same basis
in which a Gibbs state is diagonal, then
$\langle f_1|\rho_\beta|f_1\rangle=1/d$ and the fidelity-based texture
is temperature independent \citep{Wang2025}. In the present work the reference basis
is instead the fixed laboratory spin basis. Because the interacting
energy eigenvectors are coherent superpositions of local spin states,
$F_\beta$ generally depends on temperature through $p_2-p_3$. This is
a basis effect, not a departure from thermal equilibrium.

Because the textureless reference is a pure coherent state, the
maximally mixed state has $F=1/4$, $T_F=3/4$, and $R=\ln4$. Thus
``textureless'' here means closeness to the chosen uniform coherent
reference, not absence of all structure or complete statistical
uniformity.

\section{Numerical results}\label{sec:numerical}

The energy-basis populations
$r_m(t)=\langle\varphi_m|\boldsymbol{\rho}(t)|\varphi_m\rangle$
determine the usual thermodynamic and longitudinal spin observables.
We evaluate texture from $\varrho(t)$ in the local computational basis
$\{\vert\uparrow\uparrow\rangle,\vert\uparrow\downarrow\rangle,
\vert\downarrow\uparrow\rangle,\vert\downarrow\downarrow\rangle\}$.
Here $\varrho_{23}(t)$ is the coherence between
$\vert\uparrow\downarrow\rangle$ and
$\vert\downarrow\uparrow\rangle$. To compare its phase-sensitive and
phase-insensitive features, we also calculate vector chirality, purity,
and the $\ell_1$-norm of coherence,
\begin{align}
{\mathcal K}_z(t)
&=\langle S_1^xS_2^y-S_1^yS_2^x\rangle(t),\nonumber\\
{\mathcal P}(t)
&=\operatorname{Tr}\{\varrho^2(t)\},\nonumber\\
{\mathcal C}_{\ell_1}(t)
&=\sum_{m\neq n}|\varrho_{mn}(t)|.
\end{align}

The initial state is constructed by assigning Gibbs populations at
temperature $T_0$ to the four energy levels and preparing a coherence
within the $M=0$ energy sector,
\begin{align}
\boldsymbol{\rho}(0)
={}&\sum_{m=1}^{4}p_m(T_0)
\vert\varphi_m\rangle\langle\varphi_m\vert \nonumber\\
&+\sqrt{p_2(T_0)p_3(T_0)}
\left[
e^{i\phi}\vert\varphi_2\rangle\langle\varphi_3\vert
+e^{-i\phi}\vert\varphi_3\rangle\langle\varphi_2\vert
\right],
\label{eq:initial-coherent-state}
\end{align}
where $p_m(T_0)=e^{-\varepsilon_m/T_0}/Z(T_0)$. The diagonal
entries coincide with the Gibbs populations at $T_0$, while the
prepared energy-basis coherence is
\begin{equation}
\rho_{23}(0)=\sqrt{p_2(T_0)p_3(T_0)}\,e^{i\phi}.
\label{eq:initial-coherence}
\end{equation}
This coherence saturates the positivity bound and is therefore the
largest allowed coherence modulus for the prescribed Gibbs populations
in the $M=0$ block. The positivity condition fixes its maximum modulus,
not its phase. The resulting state is not Gibbsian, but provides a
controlled preparation for distinguishing population and coherence
effects.

A possible idealized preparation is a classical mixture of the polarized
states, with probabilities $p_1(T_0)$ and $p_4(T_0)$, and the pure state
$|\psi_0\rangle=[\sqrt{p_2(T_0)}|\varphi_2\rangle+
 e^{-i\phi}\sqrt{p_3(T_0)}|\varphi_3\rangle]/\sqrt{q_0}$,
with probability $q_0=p_2(T_0)+p_3(T_0)$. This mixture gives exactly
Eq.~\eqref{eq:initial-coherent-state}. In the $M=0$ branch, selective
initialization into $|\varphi_2\rangle$ followed by a phase-controlled
rotation can prepare $|\psi_0\rangle$. A differential longitudinal
control proportional to $S_1^z-S_2^z$ has a nonzero matrix element
between $|\varphi_2\rangle$ and $|\varphi_3\rangle$, allowing such
a resonant rotation at frequency $K$ in principle. This protocol assumes
selective initialization and coherent control before bath relaxation;
a unitary pulse alone cannot convert the Gibbs state into the prescribed
state, since it cannot purify its $M=0$ block.

We set $k_B=\hbar=1$ and express all energies and temperatures in units
of $J$, so that time is measured in units of $J^{-1}$. Unless stated
otherwise, the numerical results use
\begin{equation}
J=1,\qquad J_z=1,\qquad h=0.5.
\end{equation}

The energy levels $\epsilon_1$ and $\epsilon_2$ cross at $K=2$,
corresponding to $D_c=\sqrt{3}$. Thus, the values $D=0,2,4$
considered below sample both sides of this crossing. We do not
evaluate the dynamics at the exact degeneracy, where the zero-frequency
sector must be reconstructed as discussed in Sec.~\ref{sec:master}.

The environment is described by the Ohmic spectral density with
exponential cutoff
\begin{equation}
\kappa(\omega)=\eta\omega e^{-\omega/\omega_c},
\qquad \eta=0.2,\qquad \omega_c=5.
\label{eq:numerical-bath}
\end{equation}
For the parameter range considered below, the active Bohr
gaps remain within the characteristic frequency range set by $\omega_c=5$.
Changing $\eta$ or $\omega_c$ modifies the relaxation time scales
and relative channel weights, while the phase-sensitive mechanism
discussed below remains unchanged.
The same spectral density and coupling strength are assigned to all
transverse and longitudinal local bath channels; the spin-operator
matrix elements, including the transverse factor $1/2$, remain
explicit in the jump operators. With the rate convention of
Eqs.~\eqref{eq:Mast-eq} and \eqref{eq:gm-m}, the zero-frequency
coefficient is obtained from the thermal Ohmic limit,
\begin{equation}
\gamma_0
=\lim_{\omega\to0^+}\kappa(\omega)[1+n(\omega)]
=\eta T_b.
\label{eq:gamma-zero-numerical}
\end{equation}
For heating at $T_b=10$ and $\eta=0.2$, this convention gives
$\gamma_0=2$. Including the spin matrix elements, the coherence decay
rate $\Gamma_{23}=(\Lambda_2+\Lambda_3)/2$ is approximately
$1.622$ for $D=1$ and $1.467$ for $D=2$, whereas their smallest active
gaps are $0.293$ and $0.118$, respectively. For $D=1$, the transverse
channel also contains distinct positive frequencies separated by
$0.414$. These scales do not exhibit the strong separation required
for a controlled full-secular approximation. To assess nonsecular
corrections, we compared the full-secular dynamics with the
Bloch--Redfield equation, which retains cross-frequency terms, using
the same bath and neglecting the Lamb shift. For $\eta=0.2$, the
maximum difference in $R(t)$ is below $0.03$ for all cases shown,
and the first-maximum times in Table~\ref{tab:robustness} differ by
at most $0.13$. Within the shared Born--Markov approximation and the
simulated time interval, these results support the robustness of the
rugosity signatures against nonsecular corrections.
The time axis is logarithmic and covers
$10^{-2}\leqslant t\leqslant 10^2$.

Figures~\ref{fig:populations}-\ref{fig:Purity} compare heating,
$T_0=1\to T_b=10$, and cooling, $T_0=10\to T_b=1$. Since the initial
states contain prepared coherence, both protocols combine population
redistribution with coherent decay and phase evolution and are not
time reverses of one another. Figures~\ref{fig:populations} and
\ref{fig:Textures} use $D=1$; the subsequent comparisons use $D=0,2,4$.

\subsection{Microscopic origin of nonequilibrium texture}

For the preparation in Eq.~\eqref{eq:initial-coherent-state} and the
secular dynamics, the only local off-diagonal pair is
$\varrho_{23}(t)$ and its conjugate. Writing
\begin{equation}
\varrho_{23}(t)=x(t)+iy(t)
=|\varrho_{23}(t)|e^{i\theta(t)},
\label{eq:coherence-polar}
\end{equation}
one obtains
\begin{align}
{\mathcal K}_z(t)=\Im\varrho_{23}(t)=&y(t),\nonumber\\
{\mathcal C}_{\ell_1}(t)
=2|\varrho_{23}(t)|
=&2\sqrt{x^2(t)+y^2(t)},\nonumber\\
F(t)=\langle f_1|\varrho(t)|f_1\rangle
=&\frac{1+2\Re[\varrho_{23}(t)]}{4}
=\frac{1+2x(t)}{4}.
\label{eq:coherence-relations}
\end{align}
Consequently,
\begin{equation}
R(t)=-\ln\!\left[\frac{1+2x(t)}{4}\right].
\label{eq:rugosity-quadrature}
\end{equation}
For the uniform reference used in the texture formulation
\citep{Parisio2024},
\begin{equation}
F(t)=\frac{1}{4}\sum_{m,n=1}^{4}\varrho_{mn}(t),
\label{eq:coherent-grand-sum}
\end{equation}
so relative phases determine constructive or destructive
contributions. Thus $F$, ${\mathcal K}_z$, and
${\mathcal C}_{\ell_1}$ resolve, respectively, the real quadrature
entering the coherent grand sum, the imaginary quadrature, and the
modulus of the same local coherence. Rugosity is the logarithmic
mapping $R=-\ln (F)$, whereas purity and trace distance quantify global
mixedness and displacement from the reference projector.

\begin{figure}[tbp]
\centering
\includegraphics[scale=0.47]{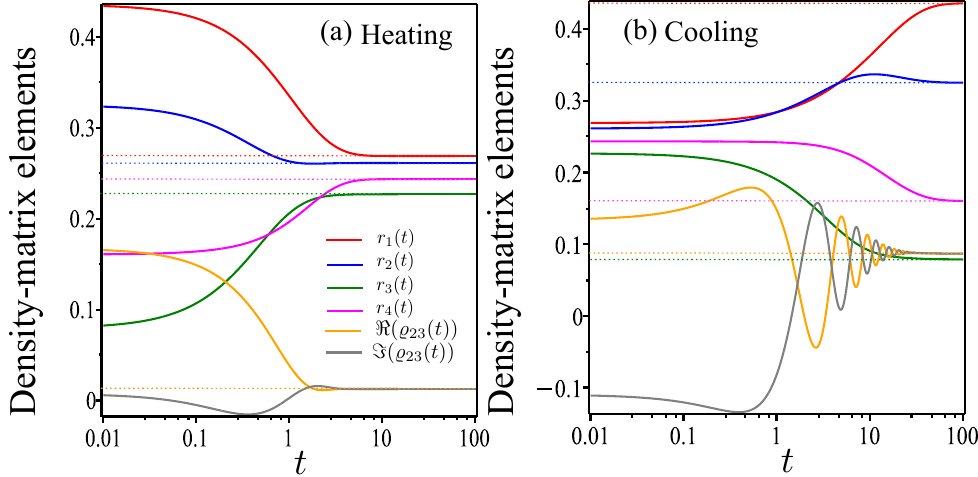}
\caption{Energy populations $r_m(t)$ and the real and imaginary parts of
the local coherence $\varrho_{23}(t)$. The latter determine the reference
overlap and vector spin chirality, respectively. Parameters are $J=J_z=1$,
$D=1$, $h=0.5$, $\eta=0.2$, and $\omega_c=5$. Panel (a): heating,
$T_0=1\to T_b=10$; panel (b): cooling, $T_0=10\to T_b=1$.}
\label{fig:populations}
\end{figure}

For the present state structure, Eq.~\eqref{eq:local-coherence-decomposition}
shows that the local coherence combines the population imbalance
$r_2-r_3$, the time-dependent
energy-basis coherence $\rho_{23}$, and the fixed DM phase
$e^{i\phi}$.
Population transfer produces the slowly varying component, whereas
the oscillatory contribution originates from the coherent evolution
of the prepared energy-basis coherence in Eq.~\eqref{eq:initial-coherence} and is progressively
damped by the bath. Such oscillations are therefore not generated
by thermalization itself; the bath damps the coherent contribution
while simultaneously redistributing the populations toward
thermal equilibrium.

In Fig.~\ref{fig:populations}, heating [panel (a)] broadens the
population distribution and reduces $r_2-r_3$. Correspondingly,
$\Re\varrho_{23}$ decreases toward a small positive value, while
$\Im\varrho_{23}$ develops a negative transient, crosses zero, and
relaxes after a small positive overshoot. During cooling [panel (b)],
population concentration increases $|r_2-r_3|$; both coherence
quadratures display stronger nonmonotonicity and damped sign changes
before reaching the low-temperature stationary values.

Texture therefore reflects the combined evolution of populations and
coherence, rather than coherence decay alone.

\subsection{Texture measures and DM-dependent dynamics}

\begin{figure}[!h]
\centering
\includegraphics[scale=0.47]{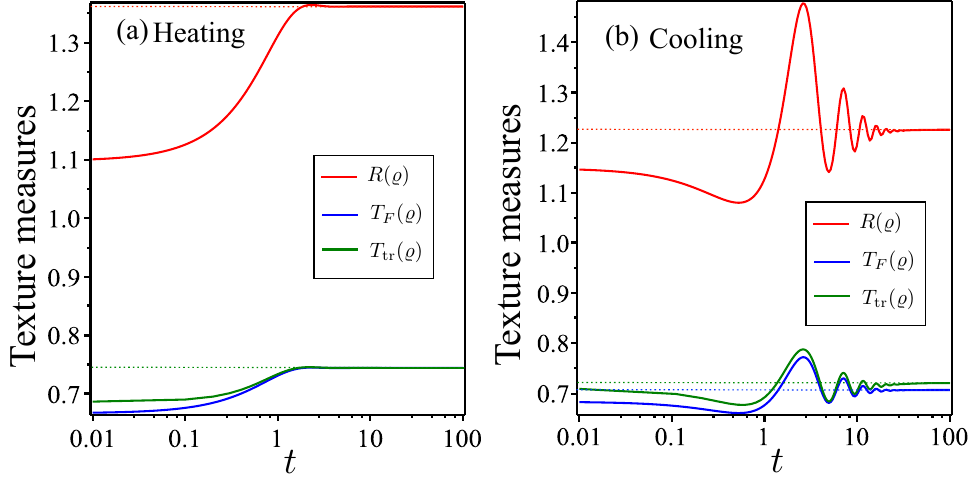}
\caption{Rugosity $R(\varrho)$, fidelity-based texture $T_F(\varrho)$, and trace-distance texture $T_{\rm tr}(\varrho)$. The quantities $R$ and $T_F$ are monotonic functions of the same overlap, $R=-\ln(1-T_F)$. Parameters are $J=J_z=1$, $D=1$, $h=0.5$, $\eta=0.2$, and $\omega_c=5$. Panel (a): heating, with a smooth approach to stationarity; panel (b): cooling, with a pronounced nonmonotonic transient.}
\label{fig:Textures}
\end{figure}

Figure~\ref{fig:Textures} compares overlap-based texture with
trace-distance distinguishability. Since $R=-\ln(1-T_F)$, $R$ and
$T_F$ have identical extrema and state ordering and differ only in
scale. By contrast, $T_{\rm tr}$ depends on the full spectrum of
$\varrho-\vert f_1\rangle\langle f_1\vert$. During heating [panel (a)],
the three curves approach stationarity predominantly smoothly, with
only a weak transient overshoot, and $T_F$ and $T_{\rm tr}$ remain
nearly coincident for the displayed parameters.
During cooling [panel (b)], all three measures exhibit a pronounced
maximum followed by damped oscillations, but $T_{\rm tr}$ weights this
transient differently from the overlap-based pair.

\begin{figure}[!h]
\centering
\includegraphics[scale=0.47]{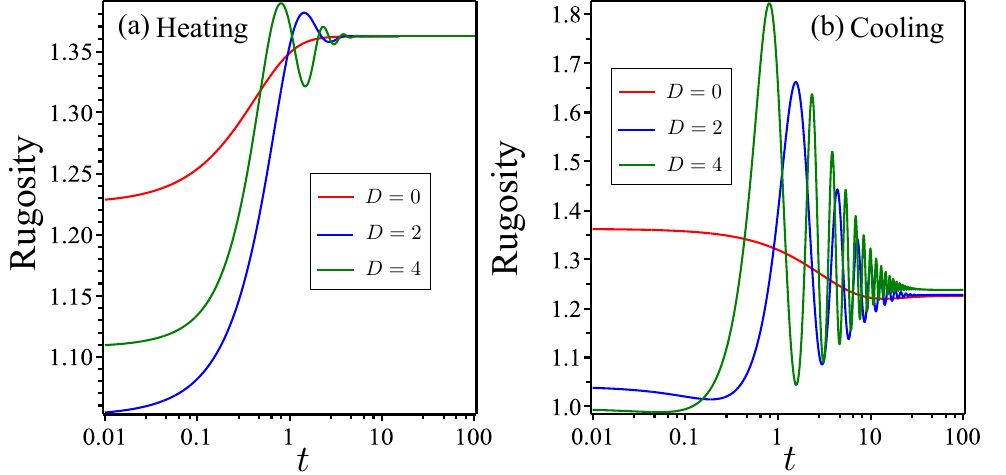}
\caption{Rugosity $R(\varrho)$ for $D=0,2,4$, showing how the DM interaction changes the dynamical departure from the uniform coherent reference. Parameters are $J=J_z=1$, $h=0.5$, $\eta=0.2$, and $\omega_c=5$. Panel (a): heating, $T_0=1\to T_b=10$; panel (b): cooling, $T_0=10\to T_b=1$.}
\label{fig:Rugosity}
\end{figure}

Figure~\ref{fig:Rugosity} shows the three interconnected effects of
$D$: it changes the splitting $K=\sqrt{J^2+D^2}$, fixes the local phase
orientation through $\phi=\arg(J+iD)$, and modifies the
dissipative rates $\gamma^\pm(\Omega_{mn})$ through the
$D$-dependent gaps. Consequently, both stationary values and transient
time scales depend on $D$.
Within the KNB realization described in Sec.~\ref{sec:model}, varying the
electric-field magnitude or orientation tunes $D$, so
Fig.~\ref{fig:Rugosity} may
be interpreted as an example of electrically controlled texture dynamics.

Combining Eqs.~\eqref{eq:coherence-polar} and
\eqref{eq:coherence-relations} gives
$F=[1+2|\varrho_{23}|\cos\theta]/4$, so rugosity responds to both the
coherence magnitude
and phase. In particular, for
$\cos\theta<0$ the coherence contribution acts to \emph{reduce} the overlap and
enhance $R$. During heating [panel (a)], the stationary values are close,
but the transients remain strongly $D$ dependent. In particular,
$D=2$ begins with the smallest rugosity. The
finite-$D$ overshoots and damped oscillations reflect evolution away
from the initial phase-sensitive alignment and become more rapid as
$D$ increases.

During cooling [panel (b)], finite $D$ produces pronounced maxima
corresponding to transiently destructive real-coherence contributions
to the reference overlap. The $D=4$ curve reaches the earliest and
largest maximum and oscillates more rapidly than the $D=2$ curve,
whereas the $D=0$ response is smooth. These peaks need not coincide
with larger coherence, greater purity, or either thermal endpoint.

\paragraph{Robustness against dissipation.}
To verify that the transient texture features are not specific to the
reference coupling $\eta=0.2$, we repeated the calculation for
$\eta=0.05$ and $0.02$. Table~\ref{tab:robustness} shows the first
finite-$D$ rugosity maximum and its occurrence time. Reducing $\eta$
by one order of magnitude preserves the maxima and leaves their
characteristic times nearly unchanged, while weaker damping generally
enhances their amplitudes and allows additional coherent oscillations
to persist. This supports the interpretation that the characteristic
texture extrema are primarily governed by the Hamiltonian phase
evolution, whereas the bath controls their damping and subsequent
relaxation. Table~\ref{tab:robustness} tests robustness within the
secular model; the Bloch--Redfield comparison separately assesses the
effect of omitted nonsecular terms.

\begin{table}[t]
\caption{Robustness of the first finite-$D$ rugosity maximum to
the system-bath coupling. $R_{\max}$ and $t_{\max}$ denote the value
and occurrence time of the first local maximum of $R(t)$, respectively.
All other parameters are those of Fig.~\ref{fig:Rugosity}.}
\label{tab:robustness}
\begin{ruledtabular}
\begin{tabular}{cccccc}
& & \multicolumn{2}{c}{Heating} & \multicolumn{2}{c}{Cooling} \\
$D$ & $\eta$ & $R_{\max}$ & $t_{\max}$ &
$R_{\max}$ & $t_{\max}$ \\
\hline
2 & 0.20 & 1.381 & 1.440 & 1.662 & 1.557 \\
2 & 0.05 & 1.438 & 1.586 & 1.835 & 1.598 \\
2 & 0.02 & 1.457 & 1.605 & 1.881 & 1.606 \\
4 & 0.20 & 1.389 & 0.808 & 1.823 & 0.806 \\
4 & 0.05 & 1.394 & 0.822 & 1.930 & 0.818 \\
4 & 0.02 & 1.393 & 0.822 & 1.955 & 0.820 \\
\end{tabular}
\end{ruledtabular}
\end{table}

\begin{figure}[!h]
\centering
\includegraphics[scale=0.47]{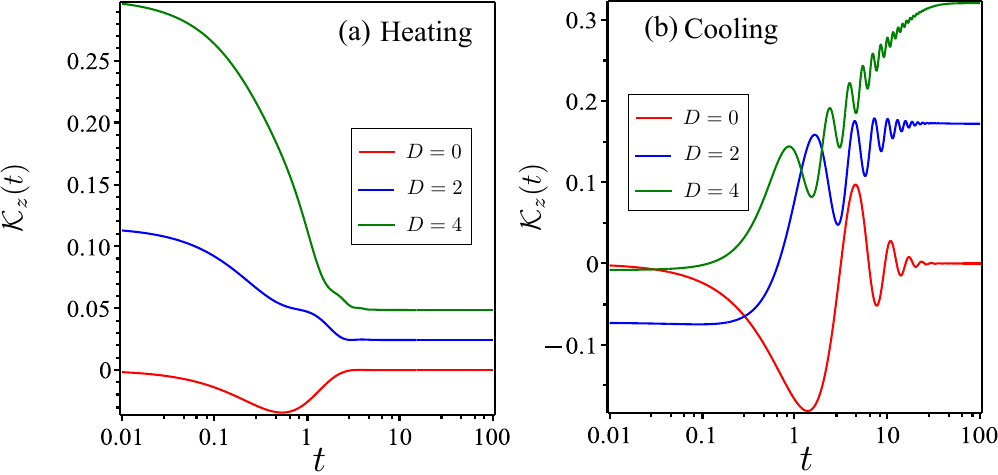}
\caption{Vector spin chirality ${\mathcal K}_z(t)=\Im(\varrho_{23}(t))$ for $D=0,2,4$. Parameters are $J=J_z=1$, $h=0.5$, $\eta=0.2$, and $\omega_c=5$. Panel (a): heating, with predominantly smooth decay for $D>0$; panel (b): cooling, with nonmonotonic oscillatory relaxation.}
\label{fig:chirality}
\end{figure}

Figure~\ref{fig:chirality} resolves the imaginary quadrature
${\mathcal K}_z=\Im(\varrho_{23})$. For $D=0$, the stationary
chirality vanishes. The negative transient during heating and the
sign-changing damped response during cooling originate from the
prepared coherence and must not be interpreted as DM-induced
equilibrium order. For $D\neq0$, the transient contribution coexists
with the thermal chirality
\begin{align}
{\mathcal K}_{z,\beta}
=\Im(\varrho_{\beta,23})
&=\frac{p_2-p_3}{2}\sin\phi\nonumber\\
&=\frac{D}{KZ}e^{-\beta J_z/4}
\sinh\!\left(\frac{\beta K}{2}\right).
\end{align}
Here $p_2-p_3$ sets the thermal amplitude and $\sin\phi=D/K$ sets the
chiral quadrature. During heating [panel (a)], the $D=2,4$ curves
decay predominantly smoothly from positive initial values toward the
smaller positive high-temperature asymptotes, with $D=4$ retaining the
larger chirality. Cooling [panel (b)] produces a strongly nonmonotonic,
oscillatory approach to the larger positive low-temperature values:
the $D=2$ curve starts negative and crosses zero, whereas the $D=4$
curve crosses zero earlier and subsequently oscillates on a positive
background. Each zero
of ${\mathcal K}_z$ marks a crossing of the real axis and need not imply
${\mathcal C}_{\ell_1}=0$. Figure~\ref{fig:Norm-l1} next isolates the
corresponding modulus.

\subsection{Texture versus coherence magnitude}

\begin{figure}[!h]
\centering
\includegraphics[scale=0.47]{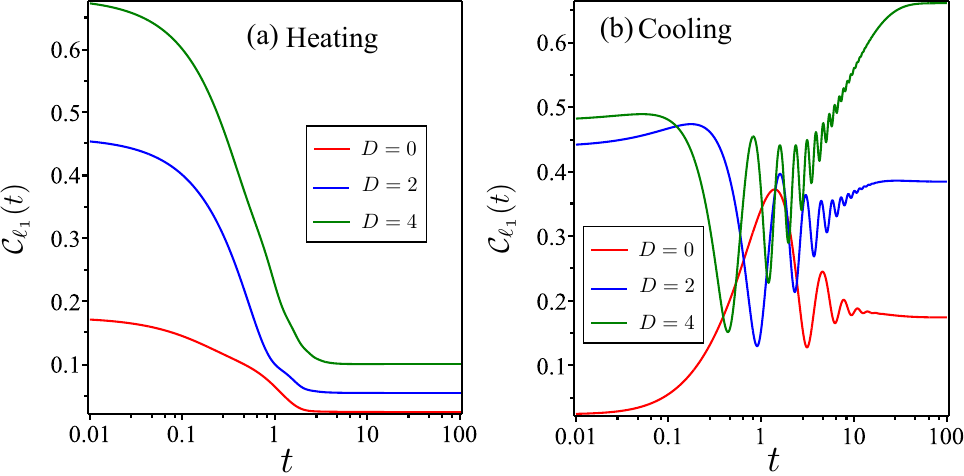}
\caption{The $\ell_1$-norm of coherence, ${\mathcal C}_{\ell_1}(t)=2|\varrho_{23}(t)|$, the coherence modulus for $D=0,2,4$, displayed independently of its phase-sensitive alignment. Parameters are $J=J_z=1$, $h=0.5$, $\eta=0.2$, and $\omega_c=5$. Panel (a): heating, $T_0=1\to T_b=10$; panel (b): cooling, $T_0=10\to T_b=1$.}
\label{fig:Norm-l1}
\end{figure}

Figure~\ref{fig:Norm-l1} displays the $\ell_1$-norm of coherence,
${\mathcal C}_{\ell_1}=2|\varrho_{23}|$, which is phase insensitive,
whereas rugosity depends on the real quadrature. Positive and negative
real coherence contribute constructively and destructively to $F$,
respectively, while purely imaginary coherence contributes to
${\mathcal C}_{\ell_1}$ and ${\mathcal K}_z$ but not directly to $F$.

During heating [panel (a)], $D=4$ has the largest initial
${\mathcal C}_{\ell_1}$, yet its rugosity is smaller than for $D=0$,
directly showing that large coherence does not imply large rugosity. During cooling
[panel (b)], the finite-$D$ curves display progressively faster
oscillations as $D$ increases. The rugosity peaks occur while both
modulus and phase evolve rapidly and therefore cannot be inferred from
${\mathcal C}_{\ell_1}$ alone.

The combined information in Figs.~\ref{fig:Rugosity},
\ref{fig:chirality}, and \ref{fig:Norm-l1} provides an operational
geometric decomposition. Since
\begin{equation}
\varrho_{23}(t)
=\frac{{\mathcal C}_{\ell_1}(t)}{2}e^{i\theta(t)},
\label{eq:coherence-modulus-phase}
\end{equation}
one has
\begin{align}
{\mathcal K}_z(t)
&=\frac{{\mathcal C}_{\ell_1}(t)}{2}\sin\theta(t),\nonumber\\
F(t)
&=\frac{1+{\mathcal C}_{\ell_1}(t)\cos\theta(t)}{4},\nonumber\\
R(t)
&=-\ln\!\left[
\frac{1+{\mathcal C}_{\ell_1}(t)\cos\theta(t)}{4}
\right].
\label{eq:geometric-decomposition}
\end{align}

Within this restricted state family, the pair
$\{R(t),{\mathcal K}_z(t)\}$ reconstructs the complete complex local
coherence:
\begin{equation}
\varrho_{23}(t)
=2e^{-R(t)}-\frac{1}{2}+i{\mathcal K}_z(t).
\label{eq:coherence-reconstruction}
\end{equation}
This reconstruction is specific to the present one-coherence family
and does not hold for a general density matrix. Within this family,
measuring rugosity and vector chirality therefore recovers the complex
local coherence without full state tomography.

\subsection{Texture versus global mixedness}

\begin{figure}[!b]
\centering
\includegraphics[scale=0.47]{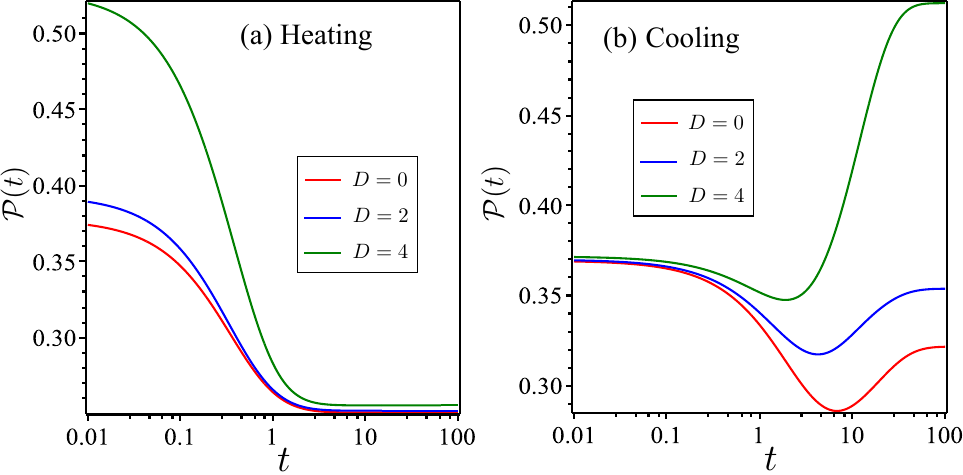}
\caption{Purity ${\mathcal P}(t)=\operatorname{Tr}\varrho^2(t)$ for $D=0,2,4$, shown to contrast phase-sensitive texture with global mixedness. Parameters are $J=J_z=1$, $h=0.5$, $\eta=0.2$, and $\omega_c=5$. Panel (a): heating, $T_0=1\to T_b=10$; panel (b): cooling, $T_0=10\to T_b=1$.}
\label{fig:Purity}
\end{figure}

Figure~\ref{fig:Purity} tests whether texture dynamics is merely a
consequence of mixedness. Purity is basis invariant and constrains the
spectrum of $\varrho$, whereas rugosity also depends on its orientation
relative to $\vert f_1\rangle$. Equal purity therefore fixes neither
quadrature of $\varrho_{23}$.

During heating [panel (a)], purity decreases almost monotonically while
rugosity exhibits strongly $D$-dependent overshoots; the initial
purity and its decay range both increase with $D$. During cooling
[panel (b)], purity first decreases through coherence loss and later
rises as populations concentrate in the colder stationary state, with
the largest final purity obtained for $D=4$. Neither behavior
determines the texture trajectory.

\section{Conclusions and outlook}

We have analyzed quantum-state texture in a thermalizing anisotropic
Heisenberg dimer with a longitudinal Dzyaloshinskii-Moriya interaction.
The global weak-coupling construction resolves the transverse and
longitudinal system-bath couplings into the Bohr-frequency channels of the
full interacting Hamiltonian. Detailed balance makes the Gibbs state
stationary, and the connected set of transition channels makes it the
relaxation state in the generic regime away from disconnected sectors and
level crossings.

The local-basis Gibbs state provides a direct physical interpretation of
the DM coupling. The combined exchange-DM magnitude enters through
$K=\sqrt{J^2+D^2}$, changing the energy splitting and the thermal population
imbalance, while the relative phase enters through
$\phi=\arg(J+iD)$, rotating the coherence between
$\vert\uparrow\downarrow\rangle$ and
$\vert\downarrow\uparrow\rangle$.

The main dynamical finding is the separation between population
thermalization and the decay and precession of the prepared energy-basis
coherence. Both enter the local coherence, so their interplay produces
DM-dependent texture overshoots and damped quadrature sign changes,
especially during cooling. These features need not track coherence
magnitude or purity, and they persist at reduced dissipation within the
secular model. The bath redistributes populations and damps the prepared
coherence; the oscillatory component originates from Hamiltonian evolution
rather than from thermalization alone.

The identities relating rugosity and fidelity-based texture express
their common reference overlap. Likewise, the reconstruction of local
coherence from rugosity and chirality follows from the restricted
one-coherence state structure. These relations provide an interpretation
and measurement route for the dynamical results rather than independent
new texture measures. Trace-distance texture depends on additional state
information and weights the transients differently.

The microscopic derivation assumes Born--Markov and full secular regimes
with uncorrelated local bath channels; the secular approximation was
checked against a nonsecular Bloch--Redfield generator in
Sec.~\ref{sec:numerical}.
Correlated reservoirs would additionally
allow interference between site-resolved transition paths. Extending the
analysis to these regimes, to driven dimers and longer chiral spin chains,
and to texture witnesses or block-texture measures
\citep{Chen2026,Chu2026} should reveal which phase-sensitive signatures
survive beyond the minimal model. Experimentally, the overlap defining
rugosity is a projection probability, while chirality is a two-spin
correlator; within the one-coherence state family considered here,
their joint accessibility provides a route for tracking the complex
local coherence without full state tomography.
A natural extension of the KNB realization is a time-dependent
electric field that dynamically controls the DM coupling.

\section*{Acknowledgements}
O.R. thanks the Brazilian agencies CNPq, CAPES, and FAPEMIG for partial financial support.
V.O. acknowledges partial financial support from CS RA MESCS (Grant No.
21AG-1C047).

\section*{Data availability}
The data that support the findings of this article are available from
the authors upon reasonable request.

\appendix
\section{Population dynamics}

To describe the level-population dynamics, we work in the energy eigenbasis
$\{\vert\varphi_{1}\rangle,\vert\varphi_{2}\rangle,
\vert\varphi_{3}\rangle,\vert\varphi_{4}\rangle\}$. The corresponding
populations are $r_{m}(t)=\rho_{mm}(t)=\langle\varphi_{m}|
\boldsymbol{\rho}(t)|\varphi_{m}\rangle$, with $m=1,\dots,4$. We collect
them into the probability vector
$\mathbf{r}(t)=(r_{1}(t),r_{2}(t),r_{3}(t),r_{4}(t))^{\top}$.

The Lindblad equation \eqref{eq:Mast-eq} yields the Pauli-type rate equation
\begin{equation}
\dot{\mathbf{r}}(t)=\mathbf{W}\mathbf{r}(t).
\end{equation}
In the generic full-secular regime in which distinct allowed
transitions have nondegenerate positive Bohr gaps, the populations
decouple from the coherences. For the numerical parameters used in
Sec.~\ref{sec:numerical}, however, $h=J_z/2$, and hence
$\Omega_{42}=\Omega_{43}=K/2$. At $D=0$, where
$K=J=J_z=1$ for the parameters considered here, one further has
$\Omega_{12}=\Omega_{42}=\Omega_{43}=1/2$. These are degeneracies of
Bohr frequencies, not of energy levels. Strictly, all rank-one
components sharing the same positive frequency must be grouped in the
corresponding spectral operator $\mathbf{S}_j^x(\omega)$ according to
the definition following Eq.~\eqref{eq:Mast-eq}. For the initial-state
family used below, the associated cross terms leave the invariant set
$\{r_m,\rho_{23}\}$ unchanged, so the equations used for the numerical
results remain valid within the grouped secular generator. In particular,
for $h=J_z/2$ the lowering components at $K/2$ are proportional to
$|\varphi_2\rangle\langle\varphi_4|$ and
$|\varphi_4\rangle\langle\varphi_3|$. Their cross anticommutator
products vanish, and their cross recycling terms require coherences
$\rho_{43}$ or $\rho_{24}$, both absent in the present state family.
The adjoint components obey the same restriction. At $D=0$, the
additional component $|\varphi_1\rangle\langle\varphi_2|$ can
combine with $|\varphi_4\rangle\langle\varphi_3|$ to transfer
$\rho_{23}$ into $\rho_{14}$, but the two site contributions cancel
because their matrix-element products have opposite signs and their
bath spectra are equal. Hence neither populations nor $\rho_{23}$
receive cross-term corrections for this preparation.

The transverse pairs are
$\{1,2\}$, $\{1,3\}$, $\{2,4\}$, and $\{3,4\}$. For the
present coupling operators, the matrix-element sums reduce to
\begin{equation}
\sum_{j=1}^{2}\sum_{\mu\in\{x,z\}}
\left|\langle\varphi_m|S_j^\mu|\varphi_n\rangle\right|^2
=
\begin{cases}
\tfrac{1}{4},&\text{transverse pairs},\\[4pt]
\tfrac{1}{2},&\{m,n\}=\{2,3\},\\[4pt]
0,&\text{otherwise}.
\end{cases}
\label{eq:matrix-element-factor}
\end{equation}
For $m\ne n$, the off-diagonal element
$W_{mn}$ is the total rate for
$\vert\varphi_n\rangle\to\vert\varphi_m\rangle$. It is given directly by
\begin{equation}
\begin{aligned}
W_{mn}
&=\frac{1}{4}
\begin{cases}
\gamma_{mn}^{-},&\varepsilon_n>\varepsilon_m,\\
\gamma_{mn}^{+},&\varepsilon_m>\varepsilon_n,
\end{cases}
&&\text{transverse pairs},\\[5pt]
W_{mn}
&=\frac{1}{2}
\begin{cases}
\gamma_{mn}^{-},&\varepsilon_n>\varepsilon_m,\\
\gamma_{mn}^{+},&\varepsilon_m>\varepsilon_n,
\end{cases}
&&\{m,n\}=\{2,3\}.
\end{aligned}
\label{eq:transition-rate}
\end{equation}
In each expression, the upper branch describes emission from the
initial higher-energy state, whereas the lower branch describes
absorption into the final higher-energy state. All other off-diagonal
rates vanish. Probability conservation fixes the diagonal elements:
\begin{equation}
W_{nn}=-\sum_{m\neq n}W_{m n}.
\label{eq:general-W}
\end{equation}
Only the five pairs
$\{1,2\}$, $\{1,3\}$, $\{2,3\}$, $\{2,4\}$, and $\{3,4\}$
are allowed by the chosen coupling operators. Their rates can vanish at
special parameter values or for spectral densities with zeros.

Combining the allowed transitions gives the explicit rate matrix
\begin{equation}
\mathbf{W}=
\begin{pmatrix}
-\Lambda_{1} & W_{12} & W_{13} & 0\\
W_{21} & -\Lambda_{2} & W_{23} & W_{24}\\
W_{31} & W_{32} & -\Lambda_{3} & W_{34}\\
0 & W_{42} & W_{43} & -\Lambda_{4}
\end{pmatrix},
\label{eq:explicit-W}
\end{equation}
where the total population outflow rates are
\begin{align}
\Lambda_{1}&=W_{21}+W_{31},\nonumber\\
\Lambda_{2}&=W_{12}+W_{32}+W_{42},\nonumber\\
\Lambda_{3}&=W_{13}+W_{23}+W_{43},\nonumber\\
\Lambda_{4}&=W_{24}+W_{34}.
\label{eq:outflow-rates}
\end{align}
The sparsity pattern of $\mathbf{W}$ reflects the separation between
population flow across magnetization sectors and mixing within the
$M=0$ subspace. Probability is conserved, because each column of
$\mathbf{W}$ sums to zero, ensuring $\sum_{m}r_{m}(t)=1$.

Provided that the dissipative channels allow population transfer among
all four energy levels, the vector $\mathbf{r}(t)$ converges in the
long-time limit to a unique stationary distribution,
\begin{equation}
\mathbf{p}=\lim_{t\to\infty}\mathbf{r}(t),\qquad p_{m}=r_{m}(\infty).
\end{equation}
The pure-dephasing coefficients do not appear because
pure dephasing does not change populations.

Detailed balance identifies this stationary distribution with the
Gibbs distribution:

\begin{equation}
\mathbf{W}\,\mathbf{p}=0,\qquad p_{m}=\frac{e^{-\beta\varepsilon_{m}}}{Z},
\end{equation}

Detailed balance makes $\mathbf{W}$ reversible. Introducing
$\mathbf{P}=\operatorname{diag}(p_{1},p_{2},p_{3},p_{4})$, the matrix
\begin{equation}
\widetilde{\mathbf{W}}
=\mathbf{P}^{-1/2}\mathbf{W}\mathbf{P}^{1/2}
\end{equation}
is real symmetric, because detailed balance gives $W_{mn}p_n=W_{nm}p_m$.
Its off-diagonal elements are
$\widetilde W_{mn}=\sqrt{W_{mn}W_{nm}}=\tfrac{1}{4}\gamma_{mn}^{+}e^{\beta\Omega_{mn}/2}$
for the transverse pairs and
$\widetilde W_{23}=\tfrac{1}{2}\gamma_{23}^{+}e^{\beta\Omega_{23}/2}$
for the longitudinal pair, with the matrix-element factors of
Eq.~\eqref{eq:matrix-element-factor}.

When the Gibbs state is the only stationary state, the eigenvalues of
$\widetilde{\mathbf W}$ are $\lambda_0=0$ and
$\lambda_\alpha<0$, with $\alpha=1,2,3$. Since
$\widetilde{\mathbf W}$ is real symmetric, its eigenvectors
${\bf u}_\alpha$ can be chosen orthonormal. Defining the orthogonal matrix
\begin{equation}
\mathbf O=({\bf u}_0,{\bf u}_1,{\bf u}_2,{\bf u}_3),\qquad
\boldsymbol{\Lambda}
=\operatorname{diag}(0,\lambda_1,\lambda_2,\lambda_3),
\end{equation}
the diagonalization reads
\begin{align}
\widetilde{\mathbf W}\mathbf O
&=\mathbf O\boldsymbol{\Lambda},\nonumber\\
\widetilde{\mathbf W}
&=\mathbf O\boldsymbol{\Lambda}\mathbf O^\top,
\qquad
{\bf u}_0=\mathbf P^{1/2}(1,1,1,1)^\top.
\label{eq:diagonalization-W}
\end{align}
The original rate matrix is therefore
\begin{equation}
\mathbf W
=\mathbf P^{1/2}\mathbf O\boldsymbol{\Lambda}
\mathbf O^\top\mathbf P^{-1/2}.
\end{equation}

The population dynamics can then be written as
\begin{align}
\mathbf{r}(t)=\mathbf{p}
+{}&\sum_{\alpha=1}^{3}e^{\lambda_{\alpha}t}
c_{\alpha}\mathbf{P}^{1/2}{\bf u}_{\alpha},
\nonumber\\
c_{\alpha}={}&{\bf u}_{\alpha}^{\top}\mathbf{P}^{-1/2}
[\mathbf{r}(0)-\mathbf{p}].
\label{eq:population-solution}
\end{align}
The relaxation eigenvalues $\lambda_\alpha$ are obtained by direct
diagonalization of the real symmetric matrix $\widetilde{\mathbf W}$.

\end{document}